\documentclass[reqno,11pt,a4]{amsart}
\usepackage{amsmath,amssymb,tabularx,setspace,color}
\usepackage{pdflscape}
\newtheorem{theorem}{Theorem}[section]

\newtheorem{definition}{Definition}
\newtheorem{remark}{Remark}[section]

\usepackage{slashbox}
\usepackage{multirow}
\usepackage[bf,small,tableposition=top]{caption}
\usepackage{amsthm}
\usepackage{lineno}
\usepackage[figuresright]{rotating}
\usepackage{enumerate}
\usepackage[T1]{fontenc}
\usepackage[utf8]{inputenc}
\usepackage{graphics}
\usepackage{graphicx}
\usepackage{lscape}
\usepackage[dvipsnames]{xcolor}
\usepackage{mathrsfs}
\usepackage{hyperref}
\usepackage{textcomp}
\usepackage{listings}
\usepackage{amsmath}
\usepackage[margin=3.5cm]{geometry}
\makeatletter
\renewcommand\section{\@startsection {section}{1}{\z@}%
                                   {-3.5ex \@plus -1ex \@minus -.2ex}%
                                   {2.3ex \@plus.2ex}%
                                   {\normalfont\large\bfseries}}
\makeatother

\newtheorem{Corollary}{Corollary}[section]

\begin{document}
	
	\doublespace
	
		\title[Test for independence in discrete set up]{ A non-parametric test for independence of time to failure and cause of failure for discrete competing risks data}
\onehalfspace
	\author[S\lowercase{reedevi} E. P., S\lowercase{udheesh} K. K \lowercase{and} I\lowercase{sha} D\lowercase{ewan}]{S\lowercase{reedevi} E. P.$^{*}$, S\lowercase{udheesh} K. K$^{**}$ \lowercase{and} I\lowercase{sha} D\lowercase{ewan}$^{***}$ \\
$^{*}$SNGS C\lowercase{ollege}, P\lowercase{attambi}, I\lowercase{ndia},\\
		$^{**}$I\lowercase{ndian} S\lowercase{tatistical} I\lowercase{nstitute}, C\lowercase{hennai}, I\lowercase{ndia},\\
		$^{***}$I\lowercase{ndian} S\lowercase{tatistical} I\lowercase{nstitute}, N\lowercase{ew} D\lowercase{elhi}, I\lowercase{ndia}.}
\doublespace
    \begin{abstract} Competing risks data with discrete lifetime comes up in practice. However, only limited literature exists for such data.  In this paper, we propose a non-parametric test based on U-statistics for testing  independence of time to failure and cause of failure of competing risks data when the lifetime is a discrete random  variable.  Asymptotic distribution of the proposed test statistic is derived. An extensive Monte Carlo simulation study is conducted  to assess the finite sample performance  of the proposed test. The flexibility of the testing procedure is  illustrated using  real data sets on oral cancer patients and drug exposed pregnancies.
\end{abstract}
\maketitle
\vspace{-0.35in}
\section{Introduction}\vspace{-0.05in}
\noindent In many survival and reliability studies the lifetime of the individuals are classified in terms of cause of failure. Competing risks models are used to analyse such data. The general approach to analyse to the  competing risks data is to  consider it as a bivariate random pair $(T,C)$, where $T$ is the failure time random variable and $C \in \{1,2,...,k\}$ is the corresponding cause of failure. It is often  assume that the  lifetime random variable $T$ is continuous.

In many situations the lifetime variable can be described only by a set of non negative integer valued random variables. For example, (a) a device can be monitored only once per time period
and the random variable of interest is the number of time
periods successfully completed  prior to its
failure (b) number of days a patient stayed in a hospital after a surgery (Nawata et al., 2009).   Discrete lifetime models are used in such situations. For the analysis and recent developments on discrete lifetime data, one   may refer to Nanda and Sengupta (2005),   Lai and Xie (2006), Yu (2007), Sudheesh and Nair (2010), Dewan and Sudheesh (2011), Sudheesh et al. (2015),  Dewan et al. (2016),  Cha  and Finkelstein (2019),  Lee et al. (2019),  Berger et al. (2020),  Heyard et al. (2020),  Schimid and Berger (2020) and Wen and Chen (2020).  Crowder (1997) showed that the identifiability issues arising in competing risks analysis (Tsiatis, 1975) can be resolved, under a certain set of conditions, when $T$ is a discrete random variable.

In discrete  competing risks, cause specific sub-distribution functions  (cumulative incidence functions) are given by
\begin{equation*}
F_j(t)=P(T \le  t, C=j),\quad j=1,2,..,k,
\end{equation*}
where the overall distribution function of lifetime variable $T$ is
\begin{equation*}
F(t)=P(T \le t)=\sum_{j=1}^{k} F_j(t).
\end{equation*}
Now the survival function $S(t)$ of $T$ is given by $S(t)=P(T>t)=1-F(t)$. The cause specific hazard rate is defined as
\begin{equation*}
\lambda_j(t)=P(T=t,C= j\vert T \ge t )=\frac{f_j(t)}{S(t-1)},\,j=1,2,...k,
\end{equation*}
where $f_j(t)=P(T=t,\,C=j)$. And the overall hazard rate function is
$$\lambda (t)=\sum_{j=1}^{k}\lambda_j(t).$$

In modelling competing risks data using the observable random pair $(T,C)$, the nature of dependence between $T$ and $C$ has vital importance. If $T$ and $C$ are independent, we have $F_j(t)=P(C=j)F(t)$  and it allows us to analyse $T$ and $C$ separately. Also, the time to failure and cause of failure are independent if and only if the cause specific hazard rate functions are proportional (Crowder, 2001). A number of tests are available in literature to test the independence of time to failure and cause of failure, when the lifetime variable is continuous. Interested readers may refer to Dewan et al. (2004), Sankaran et al. (2017) and Anjana et al. (2019) and the references therein. Crowder (1996) proposed a test for testing independence of latent failure times when they are discrete. However, testing independence of time to failure and cause of failure is not considered yet. This motivates us to develop a test based on U-statistics to test the independence of $T$ and $C$.

The article is organised as follows. In Section 2, we develop a non-parametric test based on  U-statistics for testing independence between $T$ and $C$. The asymptotic properties of the proposed test statistic are studied. We report the results of Monte Carlo simulation in Section 3 and the  illustrations based on two real data sets are presented in Section 4. In Section 5, we summarise the major conclusion of the study with a discussion on future works.
\vspace{-0.3in}
\section{Test Statistic}
\noindent Consider a competing risks situation with $k $ $(k>1)$ causes of failure. Let $(T,C)$ be the bivariate random vector, where $T$ is the discrete lifetime random variable  with support $\{1,2,\ldots,\}$ or a subset thereof and $C \in\{1,2,...,k\}$ is the cause of failure.  As we discussed earlier, $F_j(t)$ denote the sub-distribution function corresponding to cause $j$ for $j=1,2,..,k$ and $F(t)=\sum_{j=1}^{k}F_j(t)$ denote the overall distribution function of $T$. Also let $f(t)=P(T=t)$ and $\pi_j=P(C=j), \,j=1,2,\ldots,k$ denote the probability of observing a failure due to cause $j$.

Next we develop the test for independence between $T$ and $C$. Let $(T_i, C_i),$ $ i=1,2...,n$ be a random sample of size $n$  from $(T,C)$. On the basis of these observations, we are interested  to test the null hypothesis
\begin{equation*}
H_0: T~~~ \text{and}~~~ C~~~ \text{are independent}
\end{equation*}
against the alternative hypothesis
\begin{equation*}
H_1: T~~~~ \text{and}~~~ C~~~~ \text{are not independent}.
\end{equation*} If $T$ and $C$ are independent, $F_j(t)=\pi_jF(t)$.  Hence, to develop the test, we  define a departure measure $\Delta$ given by
\begin{equation}\label{delta}
\Delta=\sum_{t=1}^{\infty}\sum_{j=1}^{k}\big(F_j(t)-\pi_jF(t)\big)^{2}\frac{f(t)}{\pi_j}.
\end{equation}
It is obvious that $\Delta$ is zero  under the null hypothesis $H_0$ and  positive  under the alternative hypothesis $H_1$. The $\Delta$ in Eq.  (\ref{delta}) can be further simplified into
\begin{equation}\label{Delta}
\Delta=\sum_{j=1}^{k}\frac{1}{\pi_j}\sum_{t=1}^{\infty}F_j^{2}(t)f(t)-\sum_{t=1}^{\infty}F^{2}(t)f(t).
\end{equation}
Let $(T_1,C_1)$, $(T_2,C_2)$ and  $(T_3,C_3)$ be three independent and identical copies of $(T,C)$. Then we have
\begin{equation}\label{delta1}
P(\max(T_1,T_2)\le T_3, C_1=C_2=j)=\sum_{t=1}^{\infty}F_j^{2}(t)f(t)
\end{equation}
and
\begin{equation}\label{delta2}
P(\max(T_1,T_2)\le T_3)=\sum_{t=1}^{\infty}F^{2}(t)f(t).
\end{equation}
Substituting (\ref{delta1}) and (\ref{delta2}) in (\ref{Delta})  we obtain

\begin{eqnarray}\label{deltafinal}
  \Delta &=&\sum_{j=1}^{k}\frac{1}{\pi_j}P(\text{max}(T_1,T_2)\le T_3, C_1=C_2=j)-P(\text{max}(T_1,T_2) \le T_3)\nonumber\\&=&\sum_{j=1}^{k}\frac{1}{\pi_j}\Delta_{1j}-\Delta_2. \quad(\text{say})
\end{eqnarray}

Now, we estimate $\pi_j$  by $\widehat \pi_{j}=\frac{1}{n} \sum_{i=1}^{n} I(C_i=j)$, $j=1,\ldots,k$, where $I(A)$ denotes the indicator function of a set $A$.  Clearly, $\widehat \pi_j$ is the proportion of failures due to a particular cause $j$ and it is an unbiased and consistent estimator of $\pi_j$.

Next, we estimate each term in the numerator of the sum given in right hand side  of the Eq. (\ref{deltafinal}) by defining a suitable U-statistic. Consider the symmetric kernel $\psi_{1j}$, $j=1,2,...,k$  of degree 3 defined by
\begin{equation*}
\label{2.4}
\psi_{1j}((T_1,C_1),(T_2,C_2),(T_3,C_3))=\begin{cases}
1\, &\text{if} \,\, \max(T_1,T_2)\le T_3,C_1=j,C_2=j\\
& \,\text{or}\, \max(T_2,T_3)\le T_1,C_2=j,C_3=j\\
&\, \text{or}\, \max(T_1,T_3)\le T_2,C_1=j,C_3=j\\
0\, &\text{otherwise}\end{cases}.
\end{equation*}
U-statistics defined by
\begin{equation*}
\label{ustat}
U_{1j}={{{n}\choose{3}}}^{-1} \sum\limits_{1\le i<r<l\le n} \psi_{1j}\left((T_{i},C_{i}),(T_{r},C_{r}),(T_{l},C_{l})\right),\quad j=1,2,...,k,
\end{equation*}
 is the unbiased estimator of $P(\max(T_1,T_2)\le T_3,C_1=C_2=j)$.
 Also define a symmetric kernel $\psi_{2}$ given by
 \begin{equation*}
\label{2.4}
\psi_{2}(T_1,T_2,T_3)=\begin{cases}
1\, &\text{if} \,\, \,\, \max(T_1,T_2)\le T_3\\
& \,\text{or}\, \max(T_2,T_3)\le T_1\\
&\, \text{or}\, \max(T_1,T_3)\le T_2\\
0\, &\text{otherwise}
\end{cases}.
\end{equation*}Again,  a U-statistic defined by
\begin{equation*}
\label{ustat}
U_{2}={{{n}\choose{3}}}^{-1} \sum\limits_{1\le i<r<l\le n} \psi_2(T_{i},T_{r},T_{l}),
\end{equation*}
 is an unbiased estimator of $P(\max(T_1,T_2)\le T_3)$.
Hence the test statistic is given by
 \begin{equation}
 \label{deltahat}
 \widehat \Delta=\sum\limits_{j=1}^{k}\frac{U_{1j}}{\widehat \pi_j}-U_2.
 \end{equation}

Next, we study the limiting distribution of the test statistic $\widehat \Delta$. We use the large sample theory of U-statistics to prove the asymptotic results. Clearly, $U_{1j}$ and $U_{2}$ are  consistent estimators  of  $P(\max(T_1,T_2)\le T_3,C_1=C_2=j)$,  $j=1,2,...,k$ and $P(\max(T_1,T_2)\le T_3)$,  respectively (Lehmann, 1951).  Since $\widehat \pi_j$  is a consistent estimator of $\pi_j$, it follows  that  $\widehat \Delta$  is a consistent estimator of  $\Delta$. To find the asymptotic distribution of $\widehat \Delta$, first we find the asymptotic distribution of $\boldsymbol{U}=(U_{11},U_{12},...,U_{1k},U_2)$ and then use the linear transformation given in \eqref{deltahat} (replacing $\pi_j$ by $\widehat \pi_j$).

\par The asymptotic normality of  $U_2$ and $U_{1j}$ for $j=1,2,...,k$ follows from central limit theorem for U-statistics (Lee, 2019) and we  derive the corresponding results in the following theorems.
 \begin{theorem}As $n \rightarrow \infty$, $\sqrt{n}(U_{1j}-\Delta_{1j})$, $j=1,\ldots k$ is distributed as Gaussian with mean $0$ and variance  $9\sigma_{1j}^2$ where
 $$\sigma_{1j}^2= Var\Big(E\left(\psi_{1j}\big((T_1,C_1),(T_2,C_2),(T_3,C_3)|(T_1=T,C_1=C)\big)\right)\Big).$$ Under $H_0$, we have
 \begin{equation}\label{var11}
\sigma_{1j}^{2}=9Var\Big(\pi_j^{2}F^{2}(T)+2\pi_jI(C_1=j)\Big(F(T)f(T)+\sum_{x=T+1}^{\infty}F(x)f(x)\Big)\Big).
\end{equation}\end{theorem}
\noindent{\bf Proof:} By central limit theorem for U-statistics,  $n \rightarrow \infty$,  $\sqrt{n}(U_{1j}-\Delta_{1j})$, converges in distribution  to Gaussian random variable with mean $0$ and variance $9\sigma_{1j}^2$ where $\sigma_{1j}^2$ is given by (Lee, 2019)
 $$\sigma_{1j}^2= Var\Big(E\left(\psi_{1j}\big((T_1,C_1),(T_2,C_2),(T_3,C_3)|(T_1=T,C_1=C)\big)\right)\Big).$$  Consider
\begin{eqnarray*}
 &&\hskip -0.4in E(\psi_{1j}((T_1,C_1),(T_2,C_2),(T_3,C_3))|(T_1=t,C_1=c_1))\\ & =&P(max(t,T_2) \leq T_3,c_1=j,C_2=j)
+P(max(T_2,T_3) \leq t,C_1=j,C_2=j)\\&&\quad
+P(max(t,T_3) \leq T_2,c_1=j,C_3=j)
\\ & =&2P(max(t,T_2) \leq T_3,c_1=j,C_2=j)
+P(max(T_2,T_3) \leq t,C_1=j,C_2=j)
\end{eqnarray*}
Under $H_0$, we have
\begin{equation*}
P(max(T_2,T_3) \leq t,C_1=j,C_2=j)=\pi_j^{2}F^{2}(t)
\end{equation*}
and\vspace{-0.1in}
\begin{equation*}
P(max(t,T_2) \leq T_3,c_1=j,C_2=j)=\pi_jI(c_1=j)(F(t)f(t)+\sum_{x=t+1}^{\infty}F(x)f(x))
\end{equation*}
Hence
\begin{equation*}
E(\psi_{1j}(.))=\pi_j^{2}F^{2}(t)+2\pi_jI(c_1=j)\Big(F(t)f(t)+\sum_{x=t+1}^{\infty}F(x)f(x)\Big)
\end{equation*}
and
\begin{equation*}\label{var11}
\sigma_{1j}^{2}=9Var\left(\pi_j^{2}F^{2}(T)+2\pi_jI(C_1=j)\Big(F(T)f(T)+\sum_{x=T+1}^{\infty}F(x)f(x)\Big)\right).
\end{equation*}
\begin{theorem}As $n \rightarrow \infty$, $\sqrt{n}(U_2-\Delta_2)$ converges in distribution  to Gaussian with mean $0$ and variance  $9\sigma_{2}^2$ where
  \begin{equation}\label{var2}
    \sigma_{2}^2=9Var\Big(F^{2}(T)+2F(T)f(T)+2\sum_{x=T+1}^{\infty}F(x)f(x)\Big).
  \end{equation}
 \end{theorem}

\noindent{\bf Proof:} By central limit theorem for U-statistics,  $n \rightarrow \infty$, $\sqrt{n}(U_2-\Delta_2)$ converges in distribution  to Gaussian with mean $0$ and variance  $9\sigma_{2}^2$ where
$$\sigma_{2}^2=Var\left(E(\psi_2(T_1,T_2,T_3)|T_1)\right).$$
 Consider
\begin{equation*}
E(\psi_2(T_1,T_2,T_3)|T_1=t)=2P(max(t,T_2) \leq T_3)+P(max(T_2,T_3) \leq t).
\end{equation*}
Now
$$P(max(T_2,T_3) \leq t) =F^{2}(t)$$
and
\begin{equation*}
\begin{split}
P(max(t,T_2) \leq T_3) & =\sum_{y=1}^{\infty}P(max(t,y) \leq T_3)f(y) \\
&=\bar F(t-1)F(t)+\sum_{y=t+1}^{\infty}P(y \leq T_3)f(y).
\end{split}
\end{equation*}
Since $\sum_{y=t+1}^{\infty}P(y \leq T_3)f(y)=\sum_{x=t+1}^{\infty}F(x)f(x)-F(t)\bar F(t)$, we have
\begin{equation*}
E(\psi_2(T_1,T_2,T_3)|T_1=t)=F^{2}(t)+2F(t)f(t)+2\sum_{x=t+1}^{\infty}F(x)f(x).
\end{equation*}
Hence we obtain the  variance as specified in Eq. (\ref{var2}).

 To find the distribution of $\widehat{\Delta}$, we need to find the asymptotic covariance between (i) $U_{1j}$ and $U_{1s}$ for $j \neq s=1,2,..k$ and (ii) $U_{1j}$ and $U_2$ for $j=1,2,...k$. We denote $\sigma_{js}=Cov(U_{1j},U_{1s})$. Calculations of covariance terms are given in Appendix.
 Asymptotic covariance between  $U_{1j}$ and $U_{1s}$ for $j \neq s=1,2,..k$  are given by
\begin{equation}
\sigma_{js}=Cov(\psi_{1j}((T_1,C_1),(T_2,C_2),(T_3,C_3)),\psi_{1s}((T_1,C_1),(T_4,C_4),(T_5,C_5)) )
\end{equation}
which simplifies into
\begin{equation*}
\sigma_{js}=4\pi_j^{2}\pi_s^{2}\sum_{t=1}^{\infty}F^{2}(t)[F(t)f(t)+\sum_{x=t+1}^{\infty}F(x)f(x)]+
\pi_j^{2}\pi_s^{2}\sum_{t=1}^{\infty}F^{4}(t)f(t)-\Delta_{1j}\Delta_{1s}.
\end{equation*}
Also the asymptotic covariance between $U_{1j}$ and $U_2$ for $j=1,2,...k$ is given by
\begin{equation}\label{cov2}
  \sigma_{j2}=Cov(\psi_{1j}((T_1,C_1),(T_2,C_2),(T_3,C_3)),\psi_2(T_1,T_4,T_5)).
\end{equation}The above expression simplifies to
\begin{eqnarray}
  \sigma_{j2} & =& 4\pi_j^{2}\sum_{t=1}^{\infty}[F(t)f(t)+\sum_{x=t+1}^{\infty}F(x)f(x)]^{2}f(t)+
 4\pi_j^{2}\sum_{t=1}^{\infty}F^{2}(t)[F(t)f(t) \nonumber\\&&\nonumber\quad+\sum_{x=t+1}^{\infty}F(x)f(x) ]f(t) + \pi_j^{2}\sum_{t=1}^{\infty}F^{4}(t)f(t)] -\Delta_{1j}\Delta_2.
\end{eqnarray}
\begin{theorem}Under $H_0$, As $n \rightarrow \infty$, $\sqrt{n}(\boldsymbol{U}-E(\boldsymbol{U}))$ converges in distribution to  multivariate normal with  mean vector  $0$ and variance-covariance matrix   $\Sigma=((\sigma_{ij}))_{(k+1)\times (k+1)}$ where
\begin{equation*}
  \sigma_{ii}^2=\begin{cases}
                  \sigma_{1j}^2, & \mbox{if } i=1,\ldots,k \\
                  \sigma_2^2, & \mbox{if } i=k+1
                \end{cases}
\end{equation*}
and
 \begin{equation*}
  \sigma_{ij}^2=\begin{cases}
                  \sigma_{ij}, & \mbox{if } i=1,\ldots,k, j=1,\ldots,k \\
                  \sigma_{i2}, & \mbox{if } i=1,\ldots,k,\,j=k+1~~~ \text{and}~~~~ j=1,\ldots,k,\,i=k+1
                \end{cases}.
\end{equation*} \end{theorem}
\noindent Using Slutsky's theorem we derive the asymptotic distribution of $\widehat{\Delta}$.\vspace{-0.1in}
\begin{theorem}Under $H_0$, As $n \rightarrow \infty$, $\sqrt{n}\widehat{\Delta}$ converges in distribution to  normal with  mean zero and variance $\sigma_{0}^2=a'\Sigma a$, where $a=(\frac{1}{{\pi}_1},\ldots,\frac{1}{{\pi}_k},-1)'.$
\end{theorem}\vspace{-0.2in}
\noindent Using Theorem 2.4, we can find an asymptotic critical region of the proposed test. Let $\widehat \sigma_0^2$ be a consistent estimator of $\sigma_0^2$.   The test procedure  is to reject $H_0$ against $H_1$ at a significance level $\alpha$  if $$\sqrt{n}\widehat{\Delta}/\widehat{\sigma}_0> Z_{\alpha},$$ where $Z_{\alpha}$ is the upper $\alpha$ percentage point of the standard normal distribution.
\vspace{-0.2in}
\section{Simulation Studies}
\noindent An extensive simulation study  is carried out to assess the finite sample performance of the proposed test statistic. Since the exact variance of the test statistic when $H_0$ is true does not have a closed form expression, we use bootstrap re-sampling technique to calculate the critical values. We use the method proposed by Ebrahim and Boajana (2020) to find the critical points.  We generated 10000 bootstrap samples for this purpose and the critical values are obtained from the bootstrapped distribution of the test statistic. The empirical powers of the test for sample sizes $n=25,50,75$ and $100$ for significance levels $\alpha=0.05$ and $0.01$ are calculated to study the performance of the proposed test. We generated 1000 replicates of data and the empirical power is calculated as the proportion of significant test statistics. R software is used to carry out the simulation studies.

In continuous case, to generate lifetime data with competing risks, Dewan and Kulathinal (2007) proposed a family of distributions. Their methodology is used  here to generate discrete lifetime distributions  with competing risks. In discrete set up we propose a family with $k$ causes of failure having sub-distribution functions
\begin{equation}
F_{1}(t)=\pi_{1} F^{a}(t),\ldots, F_{k-1}(t)=\pi_{k-1} F^{a}(t),\,F_{k}(t)=F(t)-\sum_{j=1}^{k-1}\pi_{j} F^{a}(t), \nonumber
\end{equation}where $1 \le a \le 2$ and $0 \le \pi_{j} \le 1/k$ for $j=1,\ldots, k$. It can be easily verified that, when $a=1$, $T$ and $C$ are independent.   The  overall  distribution function of $T$ is given by
\begin{equation*}
  F(t)=F_k(t)+\sum_{j=1}^{k-1}\pi_{j} F^{a}(t).
\end{equation*}
Discrete sub-density functions are given by
\begin{equation}\label{familyk}
f_{1}(t)=\pi_{1} (F^{a}(t)-F^{a}(t-1)),\ldots, f_{k-1}(t)=\pi_{k-1}( F^{a}(t)-F^{a}(t-1))\nonumber
\end{equation}and
\begin{equation*}
  f_{k}(t)=f(t)-\sum_{j=1}^{k-1}\pi_{j} (F^{a}(t)-F^{a}(t-1)).
\end{equation*}The cause specific hazard  rate   $\lambda_j(t)$ and overall hazard rate $\lambda(t)$ can be easily obtained for our model using the  expressions given  above.

In  simulation study, we consider  situation with two competing risks. In this case, above  family of sub-distributions reduce to
\begin{equation}\label{family}
F_{1}(t)=\pi_{1} F^{a}(t)~~~~~~\text{and}~~~~~~~~~F_{2}(t)=F(t)-\pi_{1} F^{a}(t),
\end{equation}
where $1 \le a \le 2$ and $0 \le \pi_{1} \le 0.5$ and $F(t)$ is a  distribution function of a discrete random variable.  Here $\pi_{1}=P(C=1)$ is the proportion of failures due to cause 1. As mentioned, if $a=1$, $T$ and $C$ are independent and for all the other values of $a$ ($1<a\le2$), $H_1$ is true.  To generate  lifetime values, we employed two choices for  $F(t)$. First we choose $F(t)=1-(1-p)^{t}$, the distribution function of a geometric random variable with parameter $0<p<1$. Different choices of $p$ are attempted and we present the result for  $p=0.1,0.3,0.5$ and $0.7$.
We also consider a discrete Weibull distribution  with distribution function  $F(t)=1-((1-p)^{t})^\beta$, $0<p<1,\,\beta>0$  to calculate the empirical power of the test statistic. Various combinations of $p$ and $\beta$ are used to estimate the empirical power and  we present the results in Table 2 for $p=0.3$ and $\beta=0.8,1.2,1.5,2$. Without loss of generality, the value of $\pi_{1}$ is fixed at 0.5 in our simulation studies.

\begin{table}[h!]
	\caption{Empirical power: Geometric distribution  }
	\begin{small}
\scalebox{0.85}{
		\centering
		\begin{tabular}{|c|r|rr|rr|rr|rr|}
			\hline
			\cline{3-10}&&\multicolumn{2}{c|}{$p=0.1$}&\multicolumn{2}{c|}{$p=0.3$}&
			\multicolumn{2}{c|}{$p=0.5$}&\multicolumn{2}{c|}{$p=0.7$}
			
			\\
			\cline{3-10}$a$  & $n$
			&$\alpha=0.05$ & $\alpha=0.01$&$\alpha=0.05$&$\alpha=0.01$&$\alpha=0.05$&$\alpha=0.01$&$\alpha=0.05$&$\alpha=0.01$\\
			\hline
			\multirow{3}{*} 1& 25& 0.035& 0.028 & 0.032 &0.023 &0.040 &0.021 &0.039 & 0.022\\
			& 50 &0.040 & 0.024&0.037 & 0.018&0.045 &0.018 & 0.045 &0.017	\\
			& 75 &0.048 &0.018 &0.045 & 0.015&0.049 & 0.014& 0.048&0.013  	\\
			& 100 & 0.051& 0.011&0.049 &0.009 &0.051 &0.011 &0.052 & 0.011 	\\
			\hline
			\multirow{3}{*}{1.2}
			& 25&0.348 &0.243 & 0.352&0.276 &0.371 & 0.287& 0.361& 0.253\\
			& 50 &0.449 & 0.331&0.452 &0.329 & 0.442& 0.336&0.429 &0.324	\\
			& 75 &0.492 &0.389 &0.489 & 0.393&0.494 &0.399 &0.489 & 0.372 	\\
			& 100 &0.514 &0.465 & 0.531&0.459 &0.548 & 0.461&0.547 & 0.468 	\\
			\hline
			\multirow{3}{*}{1.5}
			& 25&0.756 &0.689 &0.769 &0.678 &0.772 &0.681 & 0.762&0.662 \\
			& 50 &0.795 &0.712 & 0.781& 0.690&0.801 &0.719 &0.799 &	0.708\\
			& 75 &0.869 &0.815 &0.872 & 0.809& 0.881&0.811 &0.877 & 0.819	\\
			& 100 &0.901 & 0.869&0.898 &0.876 &0.912 &0.872 &0.908 & 0.862 	\\
			\hline
			\multirow{3}{*}{1.8}
			& 25&0.892 & 0.863&0.886 & 0.857& 0.899&0.850 & 0.891&0.858 \\
			& 50 &0.932 &0.901 &0.929 &0.892 & 0.933&0.911 &0.922 &0.896	\\
			& 75 &0.965 &0.912 &0.959 & 0.921& 0.961&0.929 &0.963 &0.919  	\\
			& 100 &0.996 &0.982 &0.997 &0.979 &0.998 &0.980 &0.993 & 0.977 	\\
			\hline
			\multirow{3}{*}{2}
			& 25&0.943 &0.921 &0.948 &0.919 &0.951 &0.922 & 0.949&0.929 \\
			& 50 &0.998 &0.996 &0.998 &0.995 & 0.998&0.996 &0.997 &0.996	\\
			& 75 &1 &0.999 &0.999 &0.998 & 1&0.999 &1 &0.999  	\\
			& 100 &1 &1&1 &1 & 1&1 &1 &1  \\
			
			\hline
		\end{tabular}}
	\end{small}
	\label{31}%
\end{table}%

The algorithm used for finding  the empirical power function can be summarised as follows.
\begin{enumerate}
\item Generate lifetime random variable $T$ from desired distribution.
	\item  Generate competing risks data $(T_{i},C_{i})$, $i=1,2,...,n$ from the family of distributions specified in  (\ref{family})   and calculate the test statistic $\widehat\Delta$.
	\item Generate $(T^{'}_{i},C^{'}_{i})$, $i=1,2,...,n$ from the above family of distributions with $a=1$ and calculate the test statistic $\widehat\Delta$.
	\item Repeat the Step 3, B=10000 times to obtain the bootstrap distribution of  $\widehat\Delta$ and determine the 95\% and 99\% percentile points.
	\item Determine whether the test statistic obtained in Step 1, is significant or not.
	\item Repeat all the Steps 1-5, 1000 times and calculate empirical power as  proportion of significant test statistics.
\end{enumerate}


\begin{table}[h!]
	\caption{Empirical power: Weibull distribution with $p=0.3$ }
	\begin{small}
\scalebox{0.85}{
		\centering
		\begin{tabular}{|c|r|rr|rr|rr|rr|}
			\hline
			\cline{3-10}&&\multicolumn{2}{c|}{$\beta=0.8$}&\multicolumn{2}{c|}{$\beta=1.2$}&
			\multicolumn{2}{c|}{$\beta=1.5$}&\multicolumn{2}{c|}{$\beta=2$}
			
			\\
			\cline{3-10}$a$  & $n$
			&$\alpha=0.05$&$\alpha=0.01$&$\alpha=0.05$&$\alpha=0.01$&$\alpha=0.05$&$\alpha=0.01$&$\alpha=0.05$&$\alpha=0.01$\\
			\hline
			\multirow{3}{*} 1& 25& 0.033& 0.025 & 0.032 &0.023 &0.035 &0.021 &0.030 & 0.021\\
			& 50 &0.037 & 0.020&0.038 & 0.019&0.038 &0.016 & 0.041 &0.018	\\
			& 75 &0.043 &0.016 &0.042 & 0.015&0.044 & 0.012& 0.048&0.014  	\\
			& 100 & 0.048& 0.012&0.047 &0.012 &0.049 &0.009 &0.051 & 0.010 	\\
			\hline
			\multirow{3}{*}{1.2}
			& 25&0.372 &0.269 & 0.369&0.259 &0.363 & 0.243& 0.352& 0.241\\
			& 50 &0.434 & 0.343&0.429 &0.332 & 0.420& 0.336&0.423 &0.325	\\
			& 75 &0.489 &0.382 &0.481 & 0.378&0.472 &0.377 &0.466 & 0.365 	\\
			& 100 &0.525 &0.452 & 0.518&0.448&0.509 & 0.435&0.506 & 0.429 	\\
			\hline
			\multirow{3}{*}{1.5}
			& 25&0.760 &0.655 &0.751 &0.643 &0.743 &0.644 & 0.735&0.632 \\
			& 50 &0.789 &0.714 & 0.782& 0.698&0.771 &0.682 &0.763 &	0.680\\
			& 75 &0.863 &0.809 &0.857 & 0.800& 0.858&0.798 &0.843 & 0.792	\\
			& 100 &0.911 & 0.852&0.897 &0.846 &0.889 &0.841 &0.882 & 0.831 	\\
			\hline
			\multirow{3}{*}{1.8}
			& 25&0.878 & 0.849&0.869 & 0.831& 0.866&0.829 & 0.862&0.820 \\
			& 50 &0.915 &0.901 &0.908 &0.892 & 0.903&0.881 &0.891 &0.878	\\
			& 75 &0.947 &0.921 &0.939 & 0.918& 0.940 &0.905 &0.931 &0.906  	\\
			& 100 &0.989 &0.971 &0.979 &0.959 &0.972 &0.951 &0.961 & 0.952 	\\
			\hline
			\multirow{3}{*}{2}
			& 25&0.945 &0.921&0.940 &0.919 &0.931 &0.918 & 0.933&0.911\\
			& 50 &0.998 &0.997 &0.998 &0.996 & 0.997&0.994 &0.997 &0.994	\\
			& 75 &1 &1 &1 &0.999 & 1&0.999 &1 &0.999  	\\
			& 100 &1 &1&1 &1 & 1&1 &1 &1  \\
						\hline	
			\end{tabular}}
	\end{small}
	\label{31}%
\end{table}%
From Tables 1 and 2 we  see that the proposed test has archived its size for a sample size of 100. Even for small sample size ($n=25$), the proposed test has good power. When $a=2$, the power approaches  one for all choices of parameters considered here.  From Table 2, we observe  that the power decreases when $\beta$ increases. When $\beta=1$, the discrete Weibull distribution reduces to the geometric distribution and the results coincide.
\vspace{-0.1in}
\section{Data Analysis}
In this section we illustrate the proposed testing method by applying it to two  real data sets. To find  the critical values of the test statistics for a given data, one has to follow the procedure given below.
\begin{enumerate}
  \item Under the assumption of independence of $T$ and $C$ we have $$P(C=1)=P(C=2)=\ldots=P(C=k)=\frac{1}{k}.$$
  \item Based on the assumption given in Step 1,  we generate causes of failure from a discrete uniform distribution on  $\{1,\ldots,k\}$ corresponding to the  lifetime $T$.
  \item Find the test statistic using the pairs $(T,C)$ obtained in Step 2.
  \item We generate $B=1000$ bootstrap samples from the original data and repeat Steps 2 and 3 to generate sample under null hypothesis of independence.
  \item Using  these $B$ bootstrap samples we obtain the empirical distribution of  $\widehat\Delta$ and determine the 95\% and 99\% percentile point.
\end{enumerate}

 First we consider the data  on oral cancer patients given in Lee  et al. (2019).
 The data set includes the time (in years) since diagnosis to death for 700 patients over 66 years of age. The eventual death is attributed to two causes, death due to oral cancer and death due to other causes. The data is available as the supplementary material in Lee et al. (2019) in the file `exampledata.csv'. We can see that out of 700 patients, 238 individuals died due to cancer, 315 died due to other causes while 147 are still alive (or no information) at the end of the study. As our test statistic deals with complete data, we consider the censored lifetimes as lifetimes due to a third cause  and carried out the analysis.

The test statistic is calculated using the proposed method. The critical values are obtained using the procedure discussed above and the critical points are calculated as 0.0015 and 0.0016 for levels  of significance $0.05$ and $0.01$, respectively. We calculate the test statistic for the given data and we obtain $\hat\Delta$ as 0.0007. Hence we conclude that time failure and cause of failure  are independent for the oral cancer data set.
\begin{figure}[h!]
	\centering
	\includegraphics[width=0.5\linewidth]{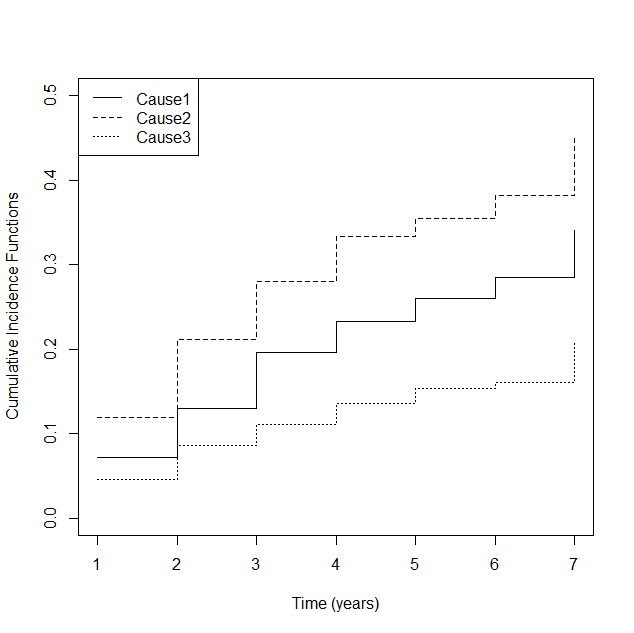}
	\caption{Cumulative incidence functions due to three causes of failure for oral cancer data set}
\end{figure}
\noindent
We plot the cumulative incidence functions of the oral cancer data set in Figure 1. From Figure 1 it is clear that, as time increases for any of the three causes, the share of their contribution to the overall distribution function does not change. Following Crowder (2002), we can conclude that time to failure and cause of failure are independent, which support our claim on oral data set using the proposed method.

In the  data discussed above, two different causes are observed for failure time. Accordingly, we analyse the data by ignoring censored lifetimes. While ignoring censored lifetimes, the data  includes lifetimes for 543 patients, out of 238 patients died due to cancer and 315 died due to other causes. The critical values are calculated using the earlier discussed method and obtained as 0.0016 and 0.0018  for critical regions of sizes $0.05$ and $0.01$ respectively. The test statistic $\hat\Delta$ is calculated as $0.0003$. It is clear that we can accept the null hypothesis that time to failure and cause of failure are independent.

\begin{figure}[h!]
	\centering
	\includegraphics[width=0.5\linewidth]{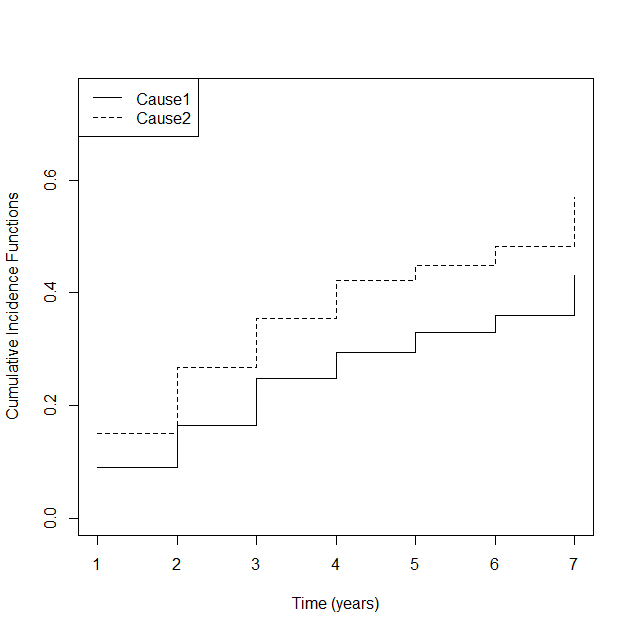}
	\caption{Cumulative incidence functions for death due to oral cancer and other causes}
		\label{Fig: 1}
\end{figure}

\begin{figure}[ht]
	\centering
	\includegraphics[width=0.5\linewidth]{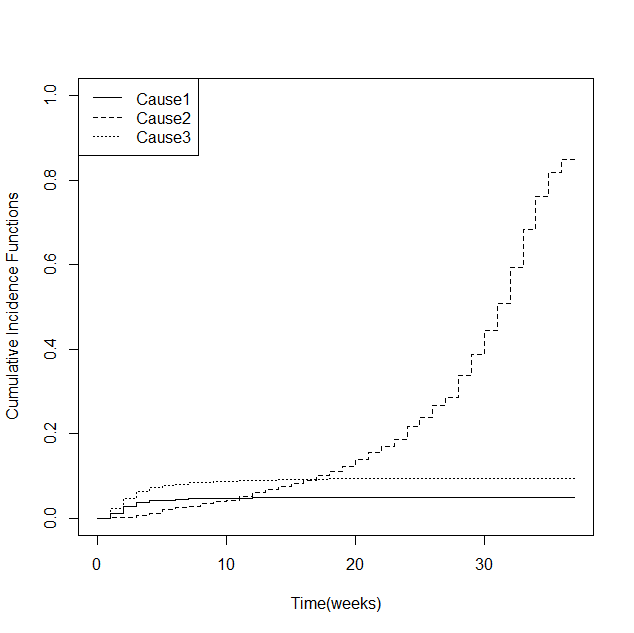}
\label{Fig:3}
	\caption{ Cumulative incidence functions of data on drug exposed pregnancies}
		
\end{figure}

We also plot the cumulative incidence functions for two causes which is given in Figure 2.
From Figure 2, we can see that the contribution of cumulative incidence functions  to the overall distribution function is not changing with respect to time.  Obviously, time to failure and cause of failure are independent for the data on 543 patients with oral cancer.

Next, we consider real life data on drug exposed pregnancies discussed in Beyersmann et al. (2011). The data set contains the information on 1186 pregnant women and the data is collected prospectively by the Teratology 	Information Service of Berlin, Germany. The pregnancy outcomes are classified as induced abortion, live birth and spontaneous abortion.  Number of weeks from conception to any of the three event is considered as the lifetime. Pregnancy outcome is known for all women. The data is available in the 'mvna' package of R- software and discussed in detail by Meister and Schaefer (2008). Out of 1186 women, 58 had induced abortion, 1016 gave live births and the remaining 112 experienced spontaneous abortion.  To examine whether the time to failure and cause of failure are independent for this data set, the critical values at rejection regions of sizes $0.05$ and $0.01$ are calculated as 0.0014 and 0.0016 respectively. The test statistic $\hat\Delta$ is obtained as 0.0359. The value of $\hat\Delta$  is grater than the critical value and we  reject the null hypothesis that time to failure and cause of failure are independent.  The plot of cumulative incidence functions for the three causes are given in Figure 3. From Figure 3, we can see that	cumulative incidence functions due to induced abortion and spontaneous abortion are dominating in earlier weeks of pregnancy, which is intuitive. Clearly, time to failure and cause of failure are dependent in this case.

\vspace{-0.2in}
\section{Concluding Remarks}
In this study, we developed a simple non-parametric test for independence of time to failure and cause of failure of competing risks data, when the lifetime is a discrete random variable.  We used U-statistics theory  to develop the test statistic and studied its asymptotic properties. A simple non-parametric bootstrap procedure is employed to calculate the critical values of the proposed test. Finite sample performance of the test is assessed through an extensive Monte Carlo simulation study by generating random variables from Geometric and discrete Weibull distributions. In both cases, the type I error of the test is approaching the chosen significance level and the test has good power.

The proposed test procedure  does not incorporate the censored observations. The tests for independence of time to failure and cause of failure for discrete lifetime data with different types of censoring can be developed.  We can also construct tests for independence  against specific dependence like positive quadrant dependent. The  proposed test  for independence can be used to study the nature of dependence between two discrete random variables where one random variable can takes finite number of distinct values while the other can take infinitely many values.


\vspace{-0.4in}
\section*{Appendix:}\vspace{-0.2in}
\section*{Covariance between $U_{1j}$ and $U_{1s}$}\vspace{-0.1in}
  \noindent Using limit theorem for U-statistics, we obtain the asymptotic covariance between  $U_{1j}$ and $U_{1s}$ for $j \neq s=1,2,...,k$  as,
\begin{eqnarray}
 \label{asycov}
\sigma_{js} &=&9 Cov(\psi_{1j}((T_1,C_1),(T_2,C_2),(T_3,C_3)),\psi_{1s}((T_1,C_1),(T_4,C_4),(T_5,C_5
 )))\nonumber \\
 &=&9\left[E(\psi_{1j}\psi_{1s})-E(\psi_{1j})E(\psi_{1s})\right].
 \end{eqnarray}\vspace{-0.15in}
Now
\begin{equation*}
E(\psi_{1j})=P(max(T_1,T_2) \leq T_3, C_1=j,C_2=j)=\Delta_{1j},\,j=1,2,...k
\end{equation*}
and
\begin{equation*}
E(\psi_2)=P(max(T_1,T_2) \leq T_3)=\Delta_2.
\end{equation*}
Now
\begin{equation*}
\begin{split}
E(\psi_{1j},\psi_{1s})& =4P(\text{max}(T_1,T_2) \leq T_3, C_1=C_2=j~~ \text{and max} (T_1,T_4) \leq T_5, C_1=C_4=s) \\
& +4P(\text{max}(T_1,T_2) \leq T_3, C_1=C_2=j~~ \text{and max} (T_4,T_5) \leq T_1, C_4=C_5=s) \\
& +P(\text{max}(T_2,T_3) \leq T_1, C_2=C_3=j~~\text{and max}(T_4,T_5) \leq T_1, C_4=C_5=s)  \\
& =4I_{11}+4I_{12}+I_{13}\quad (say).
\end{split}
\end{equation*}
 Under $H_0$, we obtain $I_{11}=0$. Consider
 \begin{equation*}
 \begin{split}
 I_{12} & =P(\text{max}(T_1,T_2) \leq T_3, C_1=C_2=j~~ \text{and max} (T_4,T_5) \leq T_1, C_4=C_5=s)\\
 & = \pi_j^{2}\pi_s^{2}\sum_{t=1}^{\infty}P(max(t,T_2) \leq T_3) P(max(T_4,T_5) \leq t) f(t) \\
 & =\pi_j^{2}\pi_s^{2}\sum_{t=1}^{\infty}F^{2}(t)P(max(t,T_2) \leq T_3) \\
 & =\pi_j^{2}\pi_s^{2}\sum_{t=1}^{\infty}F^{2}(t)\left[F(t)f(t)+\sum_{x=t+1}^{\infty}F(x)f(x)\right]. \\
 \end{split}
  \end{equation*} Now, consider 
\begin{equation*}
\begin{split}
I_{13} & =P(\text{max}(T_2,T_3) \leq T_1, C_2=C_3=j~~\text{and max}(T_4,T_5) \leq T_1, C_4=C_5=s)\\
& =\pi_j^{2}\pi_s^{2}\sum_{t=1}^{\infty}[P(max(T_2,T_3)\leq t)]^{2}f(t) \\
& =\pi_j^{2}\pi_s^{2}\sum_{t=1}^{\infty}F^{4}(t)f(t)
\\
& =\pi_j^{2}\pi_s^{2}P(max(T_1,T_2,T_2,T_3)\leq T_5).
\end{split}
\end{equation*}
Substituting the values of  $I_{12}$ and $I_{13}$ in Eq.  (\ref{asycov}) we obtain
\begin{eqnarray*}
  \sigma_{js}&=&4\pi_j^{2}\pi_s^{2}\sum_{t=1}^{\infty}F^{2}(t)\left[F(t)f(t)+\sum_{x=t+1}^{\infty}F(x)f(x)\right]\\&&\quad+
\pi_j^{2}\pi_s^{2}P(max(T_1,T_2,T_2,T_3)\leq T_5)-\Delta_{1j}\Delta_{1s}.
\end{eqnarray*}
\section*{Covariance between $U_{1j}$ and $U_{2}$}\vspace{-0.3in}
 The asymptotic covariance between $U_{1j}$ and $U_2$ for $j=1,2,...k$ is given by
\begin{eqnarray}\label{cov2}
  \sigma_{j2}&=&Cov(\psi_{1j}((T_1,C_1),(T_2,C_2),(T_3,C_3)),\psi_2(T_1,T_4,T_5))\nonumber\\
&=&E(\psi_{1j}\psi_2)-E(\psi_{1j})E(\psi_2).
\end{eqnarray}
As previously mentioned we have, $E(\psi_{1j})=\Delta_{1j}$ and $E(\psi_2)=\Delta_2$.
Now
\begin{equation*}
\begin{split}
E(\psi_{1j},\psi_{2})& =4P(\text{max}(T_1,T_2) \leq T_3, C_1=C_2=j~~ \text{and max} (T_1,T_4) \leq T_5) \\
& +4P(\text{max}(T_1,T_2) \leq T_3, C_1=C_2=j~~ \text{and max} (T_4,T_5) \leq T_1 ) \\
& +P(\text{max}(T_2,T_3) \leq T_1, C_2=C_3=j~~\text{and max}(T_4,T_5) \leq T_1)  \\
&= I_{21}+I_{22}+I_{23} \quad (say)
\end{split}
\end{equation*}
Under $H_0$, we have
\begin{equation*}
\begin{split}
I_{21} & =P(\text{max}(T_1,T_2) \leq T_3, C_1=C_2=j~~ \text{and max} (T_1,T_4) \leq T_5)\\
& = \pi_j^{2}\sum_{t=1}^{\infty}P(\text{max}(t,T_2) \leq T_3).P(\text{max}(t,T_4) \leq T_5)f(t) \\
& = \pi_j^{2}\sum_{t=1}^{\infty}\left[F(t)f(t)+\sum_{x=t+1}^{\infty}F(x)f(x)\right]^{2}f(t).
\end{split}
\end{equation*}
Now, consider
\begin{equation*}
\begin{split}
I_{22} & =P(\text{max}(T_1,T_2) \leq T_3, C_1=C_2=j~~ \text{and max} (T_4,T_5) \leq T_1 ) \\
& =\pi_j^{2}\sum_{t=1}^{\infty}P(max(t,T_2) \leq T_3)P(max(T_4,T_5) \leq t) \\
& =\pi_j^{2}\sum_{t=1}^{\infty}F^{2}(t)P(max(t,T_2) \leq T_3)f(t) \\
& =\pi_j^{2}\sum_{t=1}^{\infty}F^{2}(t)[F(t)f(t)+\sum_{x=t+1}^{\infty}F(x)f(x) ]f(t).
\end{split}
\end{equation*}
Also, we obtain 

\begin{equation*}
\begin{split}
I_{23} & =P(\text{max}(T_2,T_3) \leq T_1, C_2=C_3=j~~\text{and max}(T_4,T_5) \leq T_1) \\
& = \pi_j^{2}\sum_{t=1}^{\infty}[P(max(T_2,T_3) \leq t)]^{2}f(t) \\
& = \pi_j^{2}\sum_{t=1}^{\infty}F^{4}(t)f(t)\\
&= \pi_j^{2}P(max(T_1,T_2,T_2,T_3)\leq T_5).
\end{split}
\end{equation*}
Hence, it follows from Eq.  (\ref{cov2})
\begin{eqnarray}
  \sigma_{j2} & =& 4\pi_j^{2}\sum_{t=1}^{\infty}\left[F(t)f(t)+\sum_{x=t+1}^{\infty}F(x)f(x)\right]^{2}f(t) \nonumber\\&&\nonumber\quad+
 4\pi_j^{2}\sum_{t=1}^{\infty}F^{2}(t)\left[F(t)f(t)+\sum_{x=t+1}^{\infty}F(x)f(x)\right]f(t)\\
&&\quad + \pi_j^{2}P(max(T_1,T_2,T_2,T_3)\leq T_5) -\Delta_{1j}\Delta_2.\nonumber
\end{eqnarray}

\end{document}